\documentclass[usenatbib]{mn2e}
\usepackage{natbib,epsfig,mystyle}
\usepackage{amsmath,amssymb}

\title[SZ power spectrum from a merger-tree model]{SZ power spectrum and cluster
numbers from an extended merger-tree model}
\author[I. Dvorkin, Y. Rephaeli, M. Shimon]{Irina Dvorkin$^{1}$\thanks{E-mail:
irina@wise.tau.ac.il}, Yoel Rephaeli$^{1,2}$, Meir Shimon$^{1}$\\
$^{1}$School of Physics and Astronomy, Tel Aviv University, Tel Aviv, 69978,
Israel\\
$^{2}$Center for Astrophysics and Space Sciences, University of California,
San Diego, La Jolla, CA 92093-0424\\}

\begin{document}

\pagerange{\pageref{firstpage}--\pageref{lastpage}} \pubyear{2011}
\maketitle
\label{firstpage}

\begin{abstract}
We have recently developed an extended merger-tree model that efficiently 
follows hierarchical evolution of galaxy clusters and provides a quantitative 
description of both their dark matter and gas properties. We employed 
this diagnostic tool to calculate the thermal SZ power spectrum and cluster
number counts, accounting explicitly for uncertainties in the relevant statistical 
and intrinsic cluster properties, such as the halo mass function and the gas 
equation of state. Results of these calculations are compared with those 
obtained from a direct analytic treatment and from hydrodynamical simulations. 
We show that under certain assumptions on the gas mass fraction our results 
are consistent with the latest SPT measurement. Our approach can be particularly 
useful in predicting cluster number counts and their dependence on 
cluster and cosmological parameters.

\end{abstract}

\begin{keywords}
galaxies: clusters: general - large-scale structure of Universe - cosmic
background radiation
\end{keywords}

\section{Introduction}

The importance of the Sunyaev-Zel'dovich (SZ) effect as a valuable cosmological
probe and a usefull tool for discovery of high-redshift clusters is now firmly
established. In view of the sensitivity of the effect to the halo 
mass function (MF), in particular to the normalization of the matter power
spectrum, the SZ effect can be used to constrain various cosmological parameters
of the (standard) $\Lambda$CDM model \citep[e.g.,][and
references therein]{2001ApJ...560L.111H, 2011ARA&A..49..409A}, including the
equation of state of dark
energy \citep[e.g.,][]{2004PhRvD..70l3008W} and the total neutrino mass
\citep[e.g.,][]{2011MNRAS.412.1895S}, as well as alternative cosmological 
models \citep[e.g.,][]{2007MNRAS.380..637S}.

However, the SZ effect is also sensitive to the intracluster (IC) gas
physics, which is poorly known at high redshifts. 
The latest observational results from 
the \emph{Atacama Cosmology Telescope} \citep[ACT;
][]{2010ApJ...722.1148F} and the \emph{South Pole Telescope} \citep[SPT;
][]{2010ApJ...719.1045L,2011ApJ...736...61S,2011arXiv1111.0932R} 
underline the need for realistic modeling of IC gas properties as a prerequisite
for use of clusters as probes to determine 
cosmological parameters from SZ measurements. Indeed,
\citet{2011arXiv1111.0932R} concluded that the theoretical uncertainty in
predicting the SZ power spectrum is significantly larger than the statistical
errors. Several theoretical studies \citep{2010ApJ...725.1452S,
2011ApJ...727...94T, 
2011arXiv1106.3208E} have also demonstrated the sensitivity of the SZ effect to 
simplified modeling of IC gas, particularly its equation of state.

There are currently two approaches to modeling the SZ power 
spectrum: An 
analytical approach \citep{2002MNRAS.336.1256K} assumes universal profiles
of the dark matter (DM) density and the IC gas density (and pressure), whose
parameters follow simple scaling laws with mass and 
the cluster redshift (of observation). 
The relative contribution of clusters at different redshifts is
weighted according to an appropriate MF. The computational efficiency
of this method 
makes it feasible to explore 
a large portion of the parameter space, but it relies heavily on average scaling 
relations between the cluster observables, such as the concentration parameter,
gas temperature, 
and the cluster mass. In addition, this description does not include the
intrinsic scatter in these scaling relations which may depend on the cluster
mass and redshift.

The second approach is based on 
numerical simulations, either dynamical 
\citep[e.g.,][]{2009ApJ...702..368S, 2010ApJ...709..920S} or hydrodynamical 
\citep[e.g.,][]{2001PhRvD..63f3001S, 2001ApJ...549..681S,2005ApJ...626...12B, 
2006MNRAS.370.1309S, 2007MNRAS.378.1259R, 
2010ApJ...725...91B}, from which large catalogues of clusters are created. 
This method reproduces the great variety in the
observable properties of clusters, but the large computational costs lead to two
shortcomings: the volume of the simulation is limited, 
which leads to insufficient statistical sampling of high-mass 
($\sim 10^{15}M_{\odot}$) clusters, 
and it is highly inefficient in testing various cosmological 
and IC gas models. 
A variation on the numerical approach uses the Lagrangian perturbation theory 
(of which the Zel'dovich approximation is the first term) to generate 
DM halo catalogues with much less computational effort
\citep{2002MNRAS.331..587M, 2007MNRAS.382.1697H}. However, this technique too is
limited by the size of the simulation box.

In this work we use 
our \citep{2011MNRAS.412..665D} extended 
merger-tree model of cluster evolution 
to calculate the thermal SZ power spectrum and cluster number counts. This
method 
enables quantitative predictions of 
the outcome of numerical simulations by taking into account the intrinsic 
scatter in 
cluster properties that result from their different formation histories. Our 
model is an improvement over standard analytical methods in that it accounts 
for the full merger history and formation redshift of each cluster, and does 
not depend on pre-calibrated scaling relations. On the other hand, our approach 
can be used in studies of cosmological parameter estimation in a much more 
computationally efficient way than numerical simulations.

This paper is organized as follows. In Section \ref{sec:model} we
review our merger-tree model of cluster evolution and 
describe how it is integrated into the calculations of the 
SZ power spectrum and number 
counts. 
In Section \ref{sec:res} we present our results and discuss their implications 
in Section \ref{sec:dis}. 
We use the following cosmological parameters: $\Omega_m=0.25$, 
$\Omega_{\Lambda}=0.75$, $\Omega_b=0.045$, $H_0=73$ km/s/Mpc, $\sigma_8=0.8$.

\section{Extended merger-tree model}
\label{sec:model}

We build merger trees of DM halos using the modified \textsc{galform} algorithm
\citep{2008MNRAS.383..557P} which is based on the excursion set formalism
\citep{1993MNRAS.262..627L}.
The conditional MF used in this algorithm is calibrated to the
outcome of the Millennium Simulation \citep{2005Natur.435..629S} and is
consistent with the Sheth-Tormen MF \citep{1999MNRAS.308..119S}.
Consequently, we use this MF in all of our calculations. The more
updated MF presented in \citet{2008ApJ...688..709T} differs from
the Sheth-Tormen MF by up to $\sim 20\%$ at $z=0$, and by a higher
fraction at larger redshifts. 
The adoption of a specific (especially an analytic, theoretically-based) 
MF, introduces modeling uncertainties; these are discussed in 
Section \ref{sec:res}.

For a cluster with a given mass and at a given observation redshift 
we build a merger tree which represents its possible evolutionary track. 
We define the major merger events as those with a mass ratio $M_{>}/M_{<} < q$ 
for some $q$ whose value is to be determined, and treat all other processes of 
mass growth as continuous (relatively) slow accretion. 
We assume that during the violent major merger events 
DM is redistributed in the cluster potential well, while the minor mergers and 
accretion processes affect mainly the outskirts of the cluster but not its central 
spherical region. We assume that the scale radius $r_s$ of the halo remains
constant during slow accretion of matter onto the cluster, 
so that the radius of the halo $R$ grows during accretion but the 
interior region remains essentially unchanged. Note that we deviate here
somewhat from our original treatment \citep{2011MNRAS.412..665D} in which we
assumed that the concentration parameter, $c$, is constant during slow
accretion. 

Having built the merger tree, we start at the highest redshift with the
smallest masses and calculate the density profile of each halo in the tree. 
To conform with common practice, and for direct comparison with previous 
work, we assume a Navarro-Frenk-White \citep{1995MNRAS.275..720N} profile for
all halos at all times: 
\begin{equation}
\label{eq:nfw}
 \rho_d(x)=\frac{4\rho_s}{x(1+x)^2}
\end{equation}
where $x=r/r_s$, with $r_s$ the characteristic scale radius of the profile, 
and $\rho_s =\rho(1)$.
Starting with an initial distribution of concentration
parameters $c(M,z)$ for the earliest halos, we calculate the concentration
parameter of each successive halo in the tree from considerations of energy
conservation.
The outcome of this calculation is the concentration parameter of 
the halo we started with at the specified redshift of observation. Generating a 
large number of trees gives an estimate of the probability density function 
(PDF) for a given mass and redshift. Further details on the merger-tree model 
can be found in \citet{2011MNRAS.412..665D}. 

The next step is to model IC gas, which is assumed to constitute 
a small fraction of the cluster mass, and to
not significantly affect the evolution of the cluster. 
The gas equation of state is assumed to be polytropic, namely 
that the pressure and density are related 
by 
\begin{equation}
 P=P_0(\rho/\rho_0)^{\Gamma}
\end{equation}
with $\Gamma=1.2$.

The solution of the equation of hydrostatic equilibrium for a polytropic gas 
inside a potential well of a DM halo is \citep{2005ApJ...634..964O}:
\begin{equation}
\label{eq:rhogas}
	\rho(x)=\rho_0\left[1-\frac{B}{1+n}\left(1-\frac{\ln(1+x)}{x} \right)
\right]^{n} ,
\end{equation}
where $n=(\Gamma-1)^{-1}$, $B$ is given by:
\begin{equation}
	B=\frac{4\pi G\rho_s r_s^2\mu m_p}{k_B T_0} , 
\end{equation}
and $\mu m_p$ is the mean molecular weight. The temperature profile is given by:
\begin{equation}
\label{eq:tgas}
	T(x)=T_0\left[1-\frac{B}{1+n}\left(1-\frac{\ln(1+x)}{x} \right) \right].
\end{equation}

As a boundary condition we assume that the gas pressure at the virial radius obeys 
$P_{g}=f_gP_{d}$, where 
$f_g$ is the gas mass fraction and
$P_{d}=\rho_{d}\sigma^2$. 
We obtain $\sigma^2$, the DM (3D) velocity dispersion, by solving the Jeans 
equation (in the field of NFW-distributed DM). 
For the gas mass fraction we adopt a (nominal) mean value of $f_g=0.1$,
estimated from the data in \citet{2008ApJ...675..106B}, although we note that
these measurements refer to the mass inside 
$R_{2500}$, within which the mean density is $2500$ the background density (at 
the cluster redshift). 
While this may be considered a relatively small inner (and perhaps 
unrepresentative) region of the cluster, a similar mean value can be deduced 
from X-ray measurements within $R_{500}$ by \citet{2009ApJ...692.1033V}.
Below we explore other assumptions about $f_g$.

We also consider an alternative model for the IC gas, the $\beta-$model with
$\beta=2/3$, in which case the density is 

\begin{equation}
  \rho(x)=\frac{\rho_0}{1+x^2}
\end{equation}
and the temperature is given by the solution of the equation of hydrostatic
equilibrium with the (approximate) boundary 
condition $T\rightarrow 0$ at large radii.
Note that the non-dimensional radial coordinate is generally different from 
that in eq. (\ref{eq:nfw}).

The SZ power spectrum is computed using the halo approximation
\citep{2002MNRAS.336.1256K}:
\begin{equation}
 C_{\ell}=s(\chi)^2\int_0^{z_{max}}\frac{dV(z)}{dz}dz\int_{M_{min}}^{M_{max}}{
dM \frac { dn } { dM} |y_{\ell}(M,z)|^2}
\label{eq:cl}
\end{equation}
 
where $s(\chi)$ 
is the spectral dependence of the SZ signal given by:
\begin{equation}
  s(\chi)=\chi \frac{e^{\chi}+1}{e^{\chi}-1}-4,
\end{equation}
where $\chi=h\nu/k_BT_0$, 
$V(z)$ is the comoving volume per steradian, $dn/dM$ is the 
MF, and $y_{\ell}$ is the 2D Fourier transform of the projected 
Comptonization parameter, 

\begin{equation}
 y_{\ell}=\frac{4\pi r_s}{\ell_s^2}\int_0^c{dx x^2 \frac{sin(\ell
x/\ell_s)}{\ell x/ \ell_s} \zeta(x)}
\end{equation}
where 
$c=R_v/r_s$ is the concentration parameter ($R_v$ is the virial radius), 
$\ell_s=d_A(z)/r_s$, $d_A(z)$ is the angular diameter distance to the
cluster, and $\zeta(x)$ is the gas (normalized) pressure 
\begin{equation}
\label{eq:y}
\zeta(x)
=\frac{k_B\sigma_T}{m_e c^2}n_e(x)T_e(x) .
\end{equation}
Typical parameters are $z_{max}=2$, $M_{min}=10^{13}h^{-1}M_{\odot}$ and 
$M_{max}=10^{16}h^{-1}M_{\odot}$. 

We build one tree for each mass and redshift in equation (\ref{eq:cl}), and sum
over all the mass and redshift range to obtain the power spectrum. When we
repeat this calculation, the concentration parameter of each halo is 
slightly different, since it is effectively drawn from a probability
distribution, resulting in a different power spectrum. 
However, for a sufficiently large number of mass and redshift bins that are used
to calculate the integral in equation (\ref{eq:cl}), the calculation converges.
Below we present the results of $N_{trees}=5$ subsequent calculations for each
$\ell$. It is remarkable that the calculation converges for such a small
$N_{trees}$ in comparison with the number of trees that were needed to estimate
the concentration parameter PDF in \citet{2011MNRAS.412..665D}. This 
is due to the fact that the concentration parameter PDF changes slowly as a
function of mass and redshift, so for a dense grid of $M$ and $z$ in the
evaluation of equation (\ref{eq:cl}) we adequately sample each PDF.

We have also calculated the expected cluster number counts, following the method
of \citet{2004NewA....9..373S}. The SZ flux of a cluster is given by:

\begin{equation}
 \Delta F_{\nu}=\int{R_s(\hat{\Omega},\sigma_B)\Delta I_{\nu}(\hat{\Omega})
d\Omega} ,
\end{equation}
where 
$\hat{\Omega}$ 
is the direction on the sky, $R_s$ describes the
detector beam with beam size $\sigma_B$ and $\Delta I_{\nu}$ is the 
(spectral) intensity change,

\begin{equation}
 \Delta I_{\nu}=\frac{2(k_B T_0)^3}{(hc)^2}\cdot y \cdot g(\chi) .
\end{equation}

The spectral dependence is $g(\chi)={\chi}^4 e^{\chi} (e^{\chi}-1)^{-2}
s(\chi)$, and $T_0$ is the CMB
temperature. 
Then the expected number of clusters with flux greater than some 
threshold $\Delta \bar{F_{\nu}}$ is:

\begin{equation}
 N(\Delta \bar{F_{\nu}})=\int{\frac{dV}{dz}dz}\int_{\Delta \bar{F_{\nu}}} {B(M,
z) \frac{dN}{dM}dM} .
\end{equation}

The selection function is $B(M,z)=1$ if the flux of a given cluster is above the 
threshold; otherwise, $B(M,z)=0$. 
The calculations were carried out with \emph{Planck} HFI $143$ GHz channel, 
with a 
beam size of $\sigma_B=7.1'$, $\Delta \nu /\nu =0.33$, and 
flux sensitivity threshold 
of $\Delta \bar{F_{\nu}}=12.6$ mJy 
for $14$ month observation.

The computation time obviously depends on the number of trees we grow for each mass 
and redshift, and the number of mass and redshift bins used to approximate the 
integral in equation (\ref{eq:cl}). 
Calculating the power spectrum for different numbers of trees, and with various 
mass and redshift resolutions, we find that 
the calculation converges for $N_{trees}=5$, $N_{\Delta M}=100$ and $N_{\Delta z}=100$, 
in which case it takes $\sim 3$ minutes to calculate the power spectrum $C_{\ell}$ 
for a given $\ell$. Below we present results for $18$ values of $\ell$ between 
$\ell=100$ and $\ell=9000$.

\section{Results}
\label{sec:res}

\begin{figure}
\centering
\epsfig{file=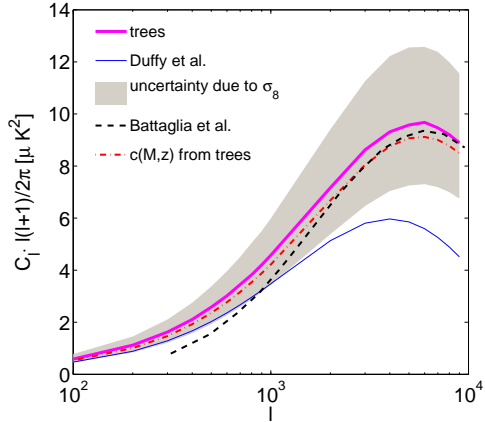, height=6cm}
\caption{SZ power spectrum 
at $\nu=153$ GHz. Results shown are from the 
standard calculation (thin blue line), merger-tree calculation (thick pink 
line), a standard calculation with $c(M,z)$ fit to the merger-tree model 
(dot-dashed red line), adiabatic hydrodynamical simulations from 
\citet{2010ApJ...725...91B} (dashed black line). The grey area shows the
uncertainty of the merger-tree calculation due to variations of $\sigma_8$.}
\label{fig:trees_vs_st}
\end{figure}

The SZ power spectrum at $\nu=153$ GHz, computed using our merger tree model, 
is shown in Figure \ref{fig:trees_vs_st}. 
For comparison we show results of a standard calculation where we use 
the scaling relation for the concentration parameter deduced by 
\citet{2008MNRAS.390L..64D} from a sample of relaxed halos. 
Also plotted is 
the result of the standard calculation where we used the average $c(M,z)$ 
relation from the merger-tree model:
\begin{equation}
\label{eq:myc}
\begin{split}
 \log c(M,& z)  =  2.417-0.1081\cdot \log \left(\frac{M}{M_{\odot}}
\right) \\
& -1.57\cdot \log (1+z)+0.1039\cdot \log \left(\frac{M}{M_{\odot}}
\right) \log (1+z) \\
& -0.2486\cdot \log^2 (1+z) .
\end{split}
\end{equation}

The normalization of the power spectrum is known to be very sensitive
to $\sigma_8$; we find the scaling $A_{SZ}\propto \sigma_8^{7.4}$, in accord
with previous studies. The grey bands in Figure \ref{fig:trees_vs_st}
correspond to 
$\sigma_8=0.8\pm 0.03$, a level of uncertainty determined from a joint analysis 
of WMAP7, BAO and SN data \citep{2011ApJS..192...18K}.

The fiducial model presented in Figure \ref{fig:trees_vs_st} results in an
overall normalization well above the level obtained from the latest SPT
measurement \citep{2011arXiv1111.0932R}: $A_{SZ}(\ell=3000)=3.65\pm 0.69$
$\mu$K$^2$. Below we will show that within the
uncertainties in the gas model, the gas mass fraction and the value of
$\sigma_8$, our 
calculation is consistent with these measurements.

To compare the above results with those from simulations, we show in Figure 
\ref{fig:trees_vs_st} the power spectrum determined 
from an adiabatic hydrodynamical simulation by \citet{2010ApJ...725...91B} 
\footnote{http://www.astro.utoronto.ca/$\sim$battaglia/ }, rescaled to 
$\nu=153$ GHz. It is very interesting 
that both the normalization and the shape of the power spectrum 
obtained in our treatment agree well with the full numerical 
calculation, in spite of the very different approach used in the simulations. 
\citet{2010ApJ...725...91B} also calculated the power spectrum in the case when 
radiative cooling, star formation, supernova 
and AGN feedback were included, and found a reduction of power at small scales 
due to expansion of the gas to larger radii. Similar results were obtained by 
\citet{2009ApJ...702..368S} and \citet{2010ApJ...709..920S}, with different 
normalizations depending on the amount of energy feedback. 
While we have not attempted to model feedback mechanisms 
in this work, the agreement of our model with numerical calculations in the 
simple adiabatic case shows that our description of the hierarchical growth of 
clusters and the intrinsic scatter in their properties is 
reasonably realistic. We stress that even though we make 
several simplifying assumptions, including that of hydrostatic equilibrium, we
still get a better agreement with simulations than in the standard approach. 
The modeling of IC gas can be further refined, as we discuss below.

There are two major differences between the merger-tree calculation and the
standard treatment with the scaling relation from \citet{2008MNRAS.390L..64D}:
the normalization in the former case is higher, and the peak shifts to higher
multipoles ($\ell \sim 6000$ vs. $\ell \sim 4000$ in the latter case). These
features can be understood in terms of the modified $c(M,z)$ relation that 
is deduced from the merger-tree algorithm. 
When we fit the average $c(M,z)$ relation from the merger-tree model to a 
general form $c=A(M/M_{pivot})^B (1+z)^C$ that is used by 
\citet{2008MNRAS.390L..64D}, we
obtain $A=11.99^{+0.42}_{-0.43}$, $B=-0.079\pm 0.005$ and $C=-0.34\pm 0.03$ for
$M_{pivot}=2\times 10^{12}h^{-1}M_{\odot}$, with $2\sigma$ confidence level
(CL), but  
this function provides a poorer fit to our results than equation 
(\ref{eq:myc}), 
whereas \citet{2008MNRAS.390L..64D} obtain for their sample of relaxed halos 
$A=9.23^{+0.17}_{-0.16}$, $B=-0.09\pm 0.009$ and $C=-0.69\pm 0.05$ ($1\sigma$
CL). Thus, the 
merger-tree model predicts a slightly higher normalization of the $c(M,z)$ 
relation, which is partly responsible for the higher normalization of the SZ 
power spectrum, and a significantly lower redshift dependence, which also 
contributes to the increase in the power spectrum normalization and shifts the 
peak to higher multipoles (since clusters at higher redshifts are more 
concentrated and have a larger contribution to the SZ power). Indeed, when we 
calculate the SZ power spectrum with the scaling relation from 
\citet{2008MNRAS.390L..64D}, but with the redshift dependence parameter 
$C=-0.34$ taken from our merger-tree model, the peak shifts to $\ell \sim 
5000$ and the normalization increases by $\sim 14\%$. An additional difference 
between the calculation performed with the $c(M,z)$ relation from equation 
(\ref{eq:myc}) and the full merger-tree calculation, appears because the 
concentration parameter has an approximately log-normal distribution function: 
there are more halos with higher-than-average $c$ than with lower-than-average.

To what extent are these predictions for the concentration parameter
reliable? There is some observational evidence that the concentration
parameter is in fact higher than predicted by N-body simulations. For example,
the concentration parameters measured by \citet{2007MNRAS.379..209S} from X-ray
observations of $34$ dynamically relaxed clusters in the redshift range
$z=0.06-0.7$ are significantly greater
than those predicted by \citet{2008MNRAS.390L..64D} for the same masses and
redshifts. In a large dataset compiled by \citet{2007MNRAS.379..190C}, which
consists of $62$ clusters at redshifts up to $z=0.89$ and masses up to at least
$\sim 2\times 10^{15}M_{\odot}$, the normalization of the concentration
parameter is found to be higher by at least $20\%$ than the results of
numerical simulations. A similar conclusion is reached by
\citet{2010MNRAS.408.2442W}, who analyzed kinematic data of $41$ clusters at
redshifts $z<0.1$. A
recent X-ray analysis of $44$ clusters in the redshift range $0.1-0.3$ by
\citet{2010A&A...524A..68E} found a normalization of $c$ consistent with
numerical simulations, although note that their constraint on the cosmological
parameters produces a rather high $\sigma_8=1.0\pm 0.2$.

Interestingly, several strong lensing
studies of massive clusters
\citep[e.g.,][]{2011MNRAS.410.1939Z} found anomalously large Einstein
radii and concentration parameters when compared with predictions from numerical
simulations. The results of the merger-tree model suggest that perhaps these
concentration parameters, although higher than the
average values, are not incompatible with the $\Lambda$CDM model.

\begin{figure}
\centering
\epsfig{file=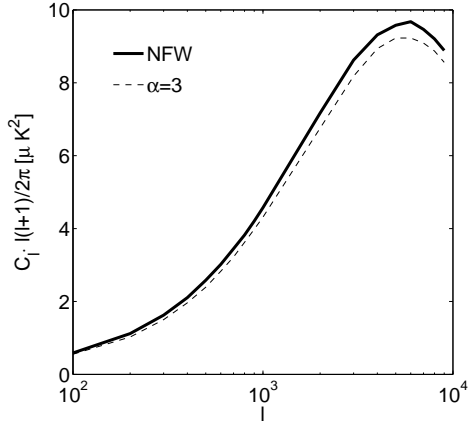, height=6cm}
\caption{SZ power spectrum for different density profiles of 
DM halos. Shown are results for an NFW profile (solid line) and 
the profile in equation (\ref{eq:mnfw}) with $\alpha=3$ (dashed line).}
\label{fig:trees_mnfw}
\end{figure}

It is important to note that the simulations of \citet{2008MNRAS.390L..64D}
span the mass range of $10^{11}-10^{15}h^{-1}M_{\odot}$, similar to other
numerical studies, while the masses that are important in the context of the SZ
effect are $10^{13}-10^{16}h^{-1}M_{\odot}$. Low-mass and low-redshift halos
outnumber high-mass halos by several orders of 
magnitude; the $c(M,z)$ relations are thus heavily weighted by the statistics 
of the low mass halos.

For example, the sample of \citet{2008MNRAS.390L..64D}
contains $1269$ sufficiently well-resolved halos at $z=0$, of which typically
less than $\sim 0.1\%$
are above $10^{14}h^{-1}M_{\odot}$ if their relative abundance approximately
follows the Sheth-Tormen MF. At $z=0.5$ only $\sim 0.03\%$ of all 
the halos are expected to have masses above $10^{14}h^{-1}M_{\odot}$.
On the other hand, the parameters $B$ and $C$ are expected to vary with both
mass and redshift \citep{2008MNRAS.387..536G, 2011MNRAS.411..584M,
2011arXiv1104.5130P}. Indeed, when we fit the average merger-tree results only
in the range $z=0-1$, $M=10^{13}-10^{14}h^{-1}M_{\odot}$ we obtain $B=-0.09$,
$C=-0.46$, closer to the values in \citet{2008MNRAS.390L..64D}. This example
demonstrates the inherent difficulty of numerical simulations to produce large
samples of halos due to computational limitations. It is possible that because
of the relatively small sample size numerical simulations overestimate the
redshift evolution of the concentration parameter which biases the predicted SZ
power spectrum. 

\begin{figure}
\centering
\epsfig{file=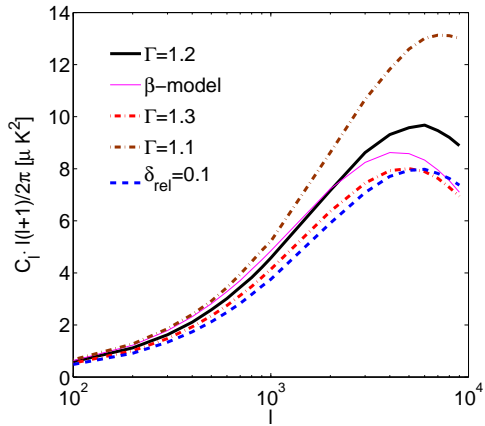, height=6cm}
\caption{SZ power spectrum for different gas models: polytropic with different
values of the adiabatic index,
non-thermal pressure component, and $\beta$-model with
$\beta=2/3$.}
\label{fig:trees_gas}
\end{figure}
Our merger-tree approach enables us to test alternative 
DM halo density profiles, which is not possible in analytical calculations 
that are based on pre-calibrated scaling relations. As an example, we 
explore the following modification of the NFW profile:

\begin{equation}
\label{eq:mnfw}
 \rho_d(r)=\frac{2^{\alpha}\rho_s}{r/r_s(1+r/r_s)^{\alpha}}
\end{equation}
for $\alpha>2$ (equations (\ref{eq:rhogas}-\ref{eq:tgas}) are replaced by the
appropriate solution for the new DM potential). Figure \ref{fig:trees_mnfw}
shows that this modification with
$\alpha=3$ has a minor effect on the amplitude of the power spectrum, 
as it affects only weakly the integrated gas pressure in the outer regions of
clusters.

\begin{figure}
\centering
\epsfig{file=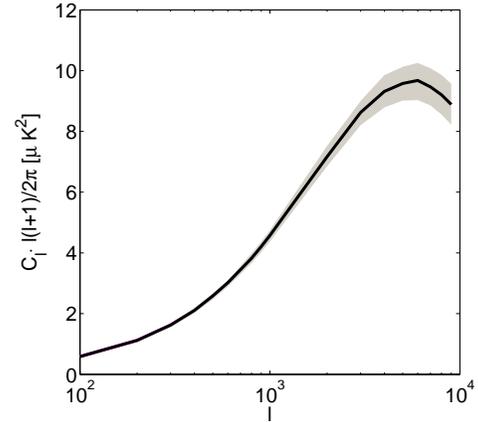, height=6cm}
\caption{SZ power spectrum with merger-tree model parameters 
in the range $q=7-13$, $\kappa=4-6$ (grey area) and the fiducial model $q=10$,
$\kappa=5$ (thick line).}
\label{fig:trees_uncer}
\end{figure}

All the calculations in this work 
were performed with the simple IC gas
model described in Section \ref{sec:model}. As was recently shown in several
studies \citep{2010ApJ...725.1452S, 2011ApJ...727...94T, 2011arXiv1106.3208E},
the strength of the SZ effect is very sensitive to the details of 
IC gas physics, for example a modest amount of non-thermal pressure can 
significantly lower the power spectrum. Here we explore the possibility of 
non-thermal pressure support, as well as slight variations in the parameters 
of the polytropic model. Figure \ref{fig:trees_gas} compares the results for
different values of the adiabatic index $\Gamma$, a $\beta-$model and a 
polytropic model with a non-thermal pressure component. For simplicity we 
assume that the non-thermal component is a constant fraction of the total 
pressure:

\begin{equation}
 P_{tot}=P_{thermal}+P_{rel}=(1+\delta_{rel})P_{thermal} ,
\end{equation}
with $\delta_{rel}=0.1$ as a representative value.

It can be seen that the SZ power spectrum is very sensitive to slight 
changes in the value of the adiabatic index away from $\Gamma=1.2$, an average 
value deduced from numerical simulations. In accord with previous studies, we find 
that a small constant fraction of non-thermal pressure obviously reduces the 
normalization of the power spectrum but does not affect its shape.

Figure \ref{fig:trees_uncer} shows the uncertainty in our calculation due 
to the merger-tree model parameters $q$, which is the maximal allowed ratio 
for major merger events, and $\kappa$, which describes the initial distance 
between two clusters that are about to merge. As expected, this uncertainty 
is smaller than $\sim 7\%$.

Our fiducial model assumes $f_g=0.1$, a constant gas mass fraction across the
whole mass and redshift range. However, the fraction of hot gas in clusters
varies with mass and redshift of observation. The mass dependence arises from 
various galactic processes including star formation, ram pressure stripping of 
gas from galaxies, and galactic winds.
While the full redshift evolution of the gas mass fraction is not known, it 
clearly reflects aspects of cluster formation and evolution, which include  
also the effects of internal processes that re-distribute cluster baryons, 
those same processes that also imprint the (related) mass dependence. The 
latter was 
deduced in several X-ray studies of low redshift clusters 
\citep{2009ApJ...692.1033V,2009ApJ...703..982G}. \citet{2009ApJ...703..982G}
derived 
the following dependence based on a combined sample of $41$ clusters at 
$z\leq 0.2$ observed by \citet{2006ApJ...640..691V}, \citet{2007A&A...474L..37A} 
and \citet{2009ApJ...693.1142S}:

\begin{equation}
f_g 
=(9.3\pm0.2)\times 10^{-2}\left(\frac{M_{500}}{2\times
10^{14}M}_{\odot} \right)^{0.21\pm 0.03} \left(\frac{h}{0.7}\right)^{-1.5} . 
\label{eq:fg_m}
\end{equation}
We note that this fit is not valid beyond $M\simeq 10^{15}M_{\odot}$.

Not knowing the explicit scaling of the gas mass fraction with redshift, 
we can only account for its mass dependence by adopting the relation 
$f_g=M_{gas,500}/M_{tot,500}=M_{gas,virial}/M_{tot,virial}$.
To avoid having 
a baryon fraction greater than the cosmic mean, which arises in 
high-mass clusters 
if the scaling relations above are strictly followed, we 
set an upper limit of $f_{cosmic}=\Omega_b/\Omega_m$.

\begin{figure}
\centering
\epsfig{file=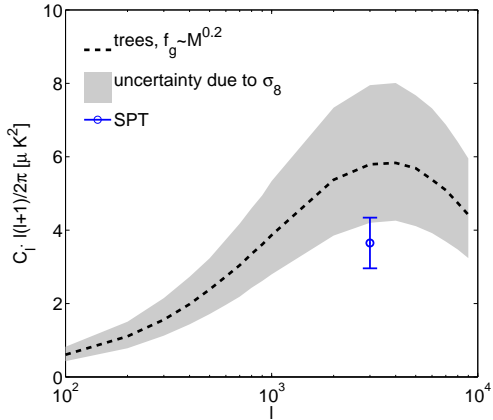, height=6cm}
\caption{SZ power spectrum calculated with the gas mass
fraction taken from equation (\ref{eq:fg_m}) and $\delta_{rel}=0.1$. Blue dot
with error bars represents the SPT measurement \citep{2011arXiv1111.0932R} at
$\ell=3000$. The grey area shows the uncertainty due to
variation in $\sigma_8$.}
\label{fig:SPT}
\end{figure}

The power spectrum, calculated with the gas mass fraction from equation
(\ref{eq:fg_m}) and assuming $\delta_{rel}=0.1$ is shown in Figure
\ref{fig:SPT}. It can be seen that our
result is consistent with the SPT measurement if we allow for an uncertainty in
$\sigma_8$ \citep{2011ApJS..192...18K}: $\sigma_8=0.8\pm 0.03$. 

While we did not account for the possible redshift evolution of
the gas mass fraction, we note 
that precise measurements of the SZ power spectrum may actually be used to
deduce, 
or significantly constrain the redshift evolution of $f_g$, especially so when 
$\sigma_8$ is more precisely determined.

\begin{figure}
\centering
\epsfig{file=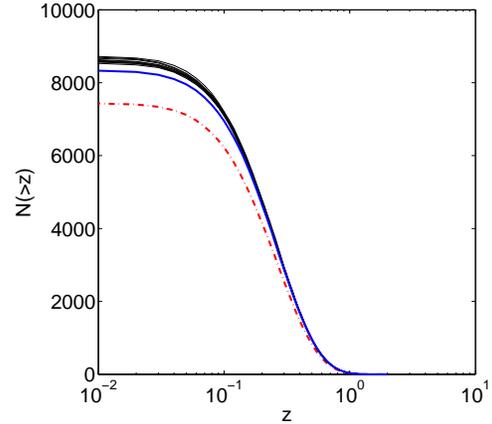, height=6cm}
\caption{SZ cluster number counts above a specified 
redshift. Shown are results of 
a standard calculation with a $c(M,z)$ relation from 
\citet{2008MNRAS.390L..64D} (red dot-dashed line), with $c(M,z)$ from a fit to
the merger-tree model (blue solid line), and the full merger-tree calculation 
(black lines).}
\label{fig:trees_nc}
\end{figure}

The expected cluster number counts are presented in Figure \ref{fig:trees_nc}.
We show the results of a standard calculation with $c(M,z)$ taken from
\citet{2008MNRAS.390L..64D} and from equation (\ref{eq:myc}), as well as the
merger-tree calculation repeated $10$ times so as to probe the distribution
functions of the cluster properties. There is $\sim 15\%$ difference between
the results of the merger-tree model and the standard calculation, 
a level of variance that could have appreciable ramifications on precise
cosmological parameter extraction
based on cluster number counts \citep{2001ApJ...560L.111H,2004PhRvD..70l3008W}. 

A major source of uncertainty in calculations of the SZ effect is the halo mass
function. It is evident
that the accuracy with which the MF can be
determined in a given numerical simulation 
is reduced 
at higher redshifts, especially at the high mass end, due to the limited volume 
of the simulation. To illustrate this point, we can estimate the expected 
number of halos at different redshifts according to the Tinker MF in 
the simulations that were used to calibrate it. For example, for $z=1.25$ the 
number of halos per unit $\ln M$ is expected to be less than $\sim 100$ for 
$M>3\cdot 10^{14} M_{\odot}$, while at $z=2.5$ there is less than one halo with 
mass around $2\cdot 10^{14} M_{\odot}$ per unit $\ln M$ in the combined 
simulation volume.

\begin{figure}
\centering
\epsfig{file=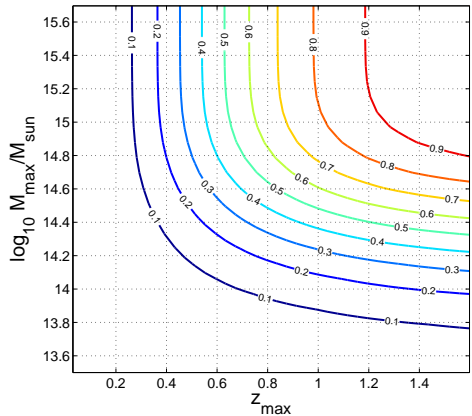, height=6cm}
\caption{The value of the integral in equation (\ref{eq:cl}) for various
$z_{max}$
(x-axis) and $M_{max}$ (y-axis) relative to the fiducial values $z_{max}=2$ and
$M_{max}=10^{16}h^{-1}M_{\odot}$, for $\ell=4000$. 
Contours 
show constant fractions of the fiducial 
value; the 
calculations were performed using the fitting formula in equation (\ref{eq:myc}).}
\label{fig:contours}
\end{figure}

It can be argued that large masses are rare at high redshifts and their
contribution to the SZ power spectrum, although uncertain, is not important.
Figure \ref{fig:contours} shows the relative contribution of different masses
and redshifts to the integral in equation (\ref{eq:cl}) in the following way: we
calculate
this expression for $\ell=4000$ with different $z_{max}$ (x-axis) and $M_{max}$
(y-axis) and
plot the result as a fraction of the fiducial value for which we took
$z_{max}=2$ and $M_{max}=10^{16}h^{-1}M_{\odot}$. 
The contours are for constant
fraction of the fiducial value\footnote{A similar plot was presented in
\citet{2011ApJ...727...94T}, except that here the variable is the maximum mass,
not the minimum mass.}. 
Two important features can be seen in
Figure
\ref{fig:contours}: (a) we have to integrate at least up to $z\sim 1.4$ to have
a reliable estimate of the effect, and (b) masses above $2\cdot 10^{14}M_{\odot}$ 
contribute roughly $\sim 50\%$ of the effect (over the whole redshift range). It
is clear that a large error in the MF even at $z\sim 1.25$ (which
only worsens at higher redshifts) is 
very detrimental to the reliability 
of the calculation.

Finally, it should be emphasized that there is an inherent limitation in 
the accuracy of cluster MF deduced from dynamical cosmological 
simulations. This stems from the appreciable impact of baryonic processes (both 
stellar and gaseous) on cluster evolution. As an example we briefly mention 
here the conclusion of \citet{2009MNRAS.394L..11S} that number densities of 
clusters in their hydrodynamical simulations deviate by $10\% -60\%$ from 
those predicted by \citet{2008ApJ...688..709T}.

\section{Discussion}
\label{sec:dis}

We have presented a new analytical method of calculating the thermal SZ power
spectrum
and cluster number 
counts. This merger-tree method 
takes into account the intrinsic scatter in the halo parameters and their 
dependence on the redshifts of formation and observation that results from 
the hierarchical evolution of clusters. In particular, our model provides a 
good statistical description of the high-mass and high-redshift halos.

We have demonstrated that our approach, which does not rely on pre-calibrated 
scaling relations, allows to explore different density profiles of 
DM halos. In addition, since the cosmological parameters can be changed in each
run of the merger-tree code and no scaling relations are inserted by hand, our
approach is more suitable for cosmological parameter estimation studies than the
standard analytic method. 

We stress that the approach presented here differs from adding scatter in the 
various observable parameters to the standard analytic calculation, since in 
this case the scaling relations and the scatter itself are inserted by hand, 
whereas here they are 
intrinsic to the model and its predictions.

Our calculations predict a higher power spectrum normalization than the
standard calculation which uses the $c(M,z)$ relation from
\citet{2008MNRAS.390L..64D} (for the same IC gas model), and a shift of the peak
to higher multipoles. In addition, the merger-tree model predicts higher
cluster number counts than the standard approach (again, for the same IC gas
model). These differences can be explained by the modified $c(M,z)$ relation
that is predicted by the merger-tree model. In particular, it has a higher
normalization and a slower evolution with redshift than the results of
\citet{2008MNRAS.390L..64D}. We argue that the higher normalization of the
concentration parameter is more compatible with observations, while the faster
redshift evolution of \citet{2008MNRAS.390L..64D} results from the fact that their 
sample is heavily weighted 
by low-mass low-redshift halos, where the dependence on redshift is expected to 
be stronger. The excellent agreement between our results and the adiabatic 
hydrodynamical simulations by \citet{2010ApJ...725...91B} 
further support our conclusions.

As anticipated, we have found the SZ power spectrum is quite sensitive to the 
assumed IC gas model. Among the parameters that can strongly influence the SZ 
power are the gas mass fraction and the amount of non-thermal pressure and
their variation with cluster mass and redshift. 
Our fiducial model, which assumes a constant $f_g=0.1$ produces a level of the
SZ power spectrum significantly higher than recent measurements by the SPT. When
we adopt a scaling of the gas mass fraction with mass 
$f_g\propto M^{0.2}$, inferred by \citet{2009ApJ...703..982G}, the (current)
SPT 
measurement falls within the range of uncertainty in $\sigma_8$ which extends to 
lower power levels due to the mass-dependent gas fraction. 
Clearly, more detailed modeling of the gas mass fraction is
needed for meaningful comparison with observations of the thermal SZ effect.
Note also that 
we adopted a class of polytropic gas models; different models can be 
readily explored in the context of our merger-tree approach. 
  
Another basic source of uncertainty is the halo MF. We have shown that the 
high-mass, high-redshift halos, for which the MF calibration is 
least certain, have a relatively large contribution to the SZ power spectrum. 
In this work we have used the Sheth-Tormen MF, but our approach can 
be extended to include other mass functions. This can be achieved by either 
calibrating the conditional MF that we use to build the merger trees 
to an up-to-date numerical simulation, in the spirit of 
\citet{2008MNRAS.383..557P}, or constructing a theoretically motivated 
conditional MF that provides a better fit to simulations.

In the calculations presented here we assumed spherical symmetry for all halos.
However, it is possible to extend our approach to account for halo triaxiality
by modeling each halo in the tree as an ellipsoid 
either with constant values of the axis ratios, or with values drawn randomly
from a probability distribution.

In this work we calculated the 
power spectrum of the SZ thermal component. 
Using our approach it is possible to calculate 
also the power due to the SZ kinematic component due to random cluster motions 
by assigning a peculiar velocity to each halo. This 
contribution 
is expected to be a small fraction of 
that from the thermal component. 
However, we cannot use our method to calculate the full kinematic 
power which also has contributions from matter outside of clusters. Therefore,
our approach is better suited for 
calculations of cluster number counts 
and cosmological parameter estimation which relies on cluster counts, since they 
depend only on the signal that originates from 
matter within clusters.

The approach presented in this work provides a computationally efficient way to
explore the uncertainties discussed above, and to gain an improved 
understanding of the influence of the evolution of galaxy clusters on their 
various observable properties.

\section*{Acknowledgments}                                                      

The authors wish to thank the \textsc{galform} team for making the code publicly
available.
Work was partly supported by the US-IL Binational Science Foundation grant 
2008452, and by a grant from the James B. Ax Family Foundation. 

\bibliographystyle{mn2e}
\bibliography{sz_ps_tree}

\begin{thebibliography}{49}
\expandafter\ifx\csname natexlab\endcsname\relax\def\natexlab#1{#1}\fi

\bibitem[{{Allen}, {Evrard} \& {Mantz}(2011){Allen}, {Evrard}, \&
  {Mantz}}]{2011ARA&A..49..409A}
{Allen} S.~W., {Evrard} A.~E., {Mantz} A.~B., 2011, \araa, 49, 409

\bibitem[{{Arnaud}, {Pointecouteau} \& {Pratt}(2007){Arnaud}, {Pointecouteau},
  \& {Pratt}}]{2007A&A...474L..37A}
{Arnaud} M., {Pointecouteau} E., {Pratt} G.~W., 2007, \aap, 474, L37

\bibitem[{{Battaglia} {et~al}\mbox{.}(2010){Battaglia}, {Bond}, {Pfrommer},
  {Sievers}, \& {Sijacki}}]{2010ApJ...725...91B}
{Battaglia} N., {Bond} J.~R., {Pfrommer} C., {Sievers} J.~L., {Sijacki} D.,
  2010, \apj, 725, 91

\bibitem[{{Bonamente} {et~al}\mbox{.}(2008){Bonamente}, {Joy}, {LaRoque},
  {Carlstrom}, {Nagai}, \& {Marrone}}]{2008ApJ...675..106B}
{Bonamente} M., {Joy} M., {LaRoque} S.~J., {Carlstrom} J.~E., {Nagai} D.,
  {Marrone} D.~P., 2008, \apj, 675, 106

\bibitem[{{Bond} {et~al}\mbox{.}(2005){Bond}, {Contaldi}, {Pen}, {Pogosyan},
  {Prunet}, {Ruetalo}, {Wadsley}, {Zhang}, {Mason}, {Myers}, {Pearson},
  {Readhead}, {Sievers}, \& {Udomprasert}}]{2005ApJ...626...12B}
{Bond} J.~R. {et~al.}, 2005, \apj, 626, 12

\bibitem[{{Comerford} \& {Natarajan}(2007)}]{2007MNRAS.379..190C}
{Comerford} J.~M., {Natarajan} P., 2007, \mnras, 379, 190

\bibitem[{{Duffy} {et~al}\mbox{.}(2008){Duffy}, {Schaye}, {Kay}, \& {Dalla
  Vecchia}}]{2008MNRAS.390L..64D}
{Duffy} A.~R., {Schaye} J., {Kay} S.~T., {Dalla Vecchia} C., 2008, \mnras, 390,
  L64

\bibitem[{{Dvorkin} \& {Rephaeli}(2011)}]{2011MNRAS.412..665D}
{Dvorkin} I., {Rephaeli} Y., 2011, \mnras, 412, 665

\bibitem[{{Efstathiou} \& {Migliaccio}(2011)}]{2011arXiv1106.3208E}
{Efstathiou} G., {Migliaccio} M., 2011, ArXiv e-prints

\bibitem[{{Ettori} {et~al}\mbox{.}(2010){Ettori}, {Gastaldello}, {Leccardi},
  {Molendi}, {Rossetti}, {Buote}, \& {Meneghetti}}]{2010A&A...524A..68E}
{Ettori} S., {Gastaldello} F., {Leccardi} A., {Molendi} S., {Rossetti} M.,
  {Buote} D., {Meneghetti} M., 2010, \aap, 524, A68+

\bibitem[{{Fowler} {et~al}\mbox{.}(2010){Fowler}, {Acquaviva}, {Ade},
  {Aguirre}, {Amiri}, {Appel}, {Barrientos}, {Battistelli}, {Bond}, {Brown},
  {Burger}, {Chervenak}, {Das}, {Devlin}, {Dicker}, {Doriese}, {Dunkley},
  {D{\"u}nner}, {Essinger-Hileman}, {Fisher}, {Hajian}, {Halpern},
  {Hasselfield}, {Hern{\'a}ndez-Monteagudo}, {Hilton}, {Hilton}, {Hincks},
  {Hlozek}, {Huffenberger}, {Hughes}, {Hughes}, {Infante}, {Irwin}, {Jimenez},
  {Juin}, {Kaul}, {Klein}, {Kosowsky}, {Lau}, {Limon}, {Lin}, {Lupton},
  {Marriage}, {Marsden}, {Martocci}, {Mauskopf}, {Menanteau}, {Moodley},
  {Moseley}, {Netterfield}, {Niemack}, {Nolta}, {Page}, {Parker}, {Partridge},
  {Quintana}, {Reid}, {Sehgal}, {Sievers}, {Spergel}, {Staggs}, {Swetz},
  {Switzer}, {Thornton}, {Trac}, {Tucker}, {Verde}, {Warne}, {Wilson},
  {Wollack}, \& {Zhao}}]{2010ApJ...722.1148F}
{Fowler} J.~W. {et~al.}, 2010, \apj, 722, 1148

\bibitem[{{Gao} {et~al}\mbox{.}(2008){Gao}, {Navarro}, {Cole}, {Frenk},
  {White}, {Springel}, {Jenkins}, \& {Neto}}]{2008MNRAS.387..536G}
{Gao} L., {Navarro} J.~F., {Cole} S., {Frenk} C.~S., {White} S.~D.~M.,
  {Springel} V., {Jenkins} A., {Neto} A.~F., 2008, \mnras, 387, 536

\bibitem[{{Giodini} {et~al}\mbox{.}(2009){Giodini}, {Pierini}, {Finoguenov},
  {Pratt}, {Boehringer}, {Leauthaud}, {Guzzo}, {Aussel}, {Bolzonella}, {Capak},
  {Elvis}, {Hasinger}, {Ilbert}, {Kartaltepe}, {Koekemoer}, {Lilly}, {Massey},
  {McCracken}, {Rhodes}, {Salvato}, {Sanders}, {Scoville}, {Sasaki}, {Smolcic},
  {Taniguchi}, {Thompson}, \& {the COSMOS Collaboration}}]{2009ApJ...703..982G}
{Giodini} S. {et~al.}, 2009, \apj, 703, 982

\bibitem[{{Holder}, {Haiman} \& {Mohr}(2001){Holder}, {Haiman}, \&
  {Mohr}}]{2001ApJ...560L.111H}
{Holder} G., {Haiman} Z., {Mohr} J.~J., 2001, \apjl, 560, L111

\bibitem[{{Holder}, {McCarthy} \& {Babul}(2007){Holder}, {McCarthy}, \&
  {Babul}}]{2007MNRAS.382.1697H}
{Holder} G.~P., {McCarthy} I.~G., {Babul} A., 2007, \mnras, 382, 1697

\bibitem[{{Komatsu} \& {Seljak}(2002)}]{2002MNRAS.336.1256K}
{Komatsu} E., {Seljak} U., 2002, \mnras, 336, 1256

\bibitem[{{Komatsu} {et~al}\mbox{.}(2011){Komatsu}, {Smith}, {Dunkley},
  {Bennett}, {Gold}, {Hinshaw}, {Jarosik}, {Larson}, {Nolta}, {Page},
  {Spergel}, {Halpern}, {Hill}, {Kogut}, {Limon}, {Meyer}, {Odegard}, {Tucker},
  {Weiland}, {Wollack}, \& {Wright}}]{2011ApJS..192...18K}
{Komatsu} E. {et~al.}, 2011, \apjs, 192, 18

\bibitem[{{Lacey} \& {Cole}(1993)}]{1993MNRAS.262..627L}
{Lacey} C., {Cole} S., 1993, \mnras, 262, 627

\bibitem[{{Lueker} {et~al}\mbox{.}(2010){Lueker}, {Reichardt}, {Schaffer},
  {Zahn}, {Ade}, {Aird}, {Benson}, {Bleem}, {Carlstrom}, {Chang}, {Cho},
  {Crawford}, {Crites}, {de Haan}, {Dobbs}, {George}, {Hall}, {Halverson},
  {Holder}, {Holzapfel}, {Hrubes}, {Joy}, {Keisler}, {Knox}, {Lee}, {Leitch},
  {McMahon}, {Mehl}, {Meyer}, {Mohr}, {Montroy}, {Padin}, {Plagge}, {Pryke},
  {Ruhl}, {Shaw}, {Shirokoff}, {Spieler}, {Stalder}, {Staniszewski}, {Stark},
  {Vanderlinde}, {Vieira}, \& {Williamson}}]{2010ApJ...719.1045L}
{Lueker} M. {et~al.}, 2010, \apj, 719, 1045

\bibitem[{{Monaco}, {Theuns} \& {Taffoni}(2002){Monaco}, {Theuns}, \&
  {Taffoni}}]{2002MNRAS.331..587M}
{Monaco} P., {Theuns} T., {Taffoni} G., 2002, \mnras, 331, 587

\bibitem[{{Mu{\~n}oz-Cuartas} {et~al}\mbox{.}(2011){Mu{\~n}oz-Cuartas},
  {Macci{\`o}}, {Gottl{\"o}ber}, \& {Dutton}}]{2011MNRAS.411..584M}
{Mu{\~n}oz-Cuartas} J.~C., {Macci{\`o}} A.~V., {Gottl{\"o}ber} S., {Dutton}
  A.~A., 2011, \mnras, 411, 584

\bibitem[{{Navarro}, {Frenk} \& {White}(1995){Navarro}, {Frenk}, \&
  {White}}]{1995MNRAS.275..720N}
{Navarro} J.~F., {Frenk} C.~S., {White} S.~D.~M., 1995, \mnras, 275, 720

\bibitem[{{Ostriker}, {Bode} \& {Babul}(2005){Ostriker}, {Bode}, \&
  {Babul}}]{2005ApJ...634..964O}
{Ostriker} J.~P., {Bode} P., {Babul} A., 2005, \apj, 634, 964

\bibitem[{{Parkinson}, {Cole} \& {Helly}(2008){Parkinson}, {Cole}, \&
  {Helly}}]{2008MNRAS.383..557P}
{Parkinson} H., {Cole} S., {Helly} J., 2008, \mnras, 383, 557

\bibitem[{{Prada} {et~al}\mbox{.}(2011){Prada}, {Klypin}, {Cuesta},
  {Betancort-Rijo}, \& {Primack}}]{2011arXiv1104.5130P}
{Prada} F., {Klypin} A.~A., {Cuesta} A.~J., {Betancort-Rijo} J.~E., {Primack}
  J., 2011, ArXiv e-prints

\bibitem[{{Reichardt} {et~al}\mbox{.}(2011){Reichardt}, {Shaw}, {Zahn}, {Aird},
  {Benson}, {Bleem}, {Carlstrom}, {Chang}, {Cho}, {Crawford}, {Crites}, {de
  Haan}, {Dobbs}, {Dudley}, {George}, {Halverson}, {Holder}, {Holzapfel},
  {Hoover}, {Hou}, {Hrubes}, {Joy}, {Keisler}, {Knox}, {Lee}, {Leitch},
  {Lueker}, {Luong-Van}, {McMahon}, {Mehl}, {Meyer}, {Millea}, {Mohr},
  {Montroy}, {Natoli}, {Padin}, {Plagge}, {Pryke}, {Ruhl}, {Schaffer},
  {Shirokoff}, {Spieler}, {Staniszewski}, {Stark}, {Story}, {van Engelen},
  {Vanderlinde}, {Vieira}, \& {Williamson}}]{2011arXiv1111.0932R}
{Reichardt} C.~L. {et~al.}, 2011, ArXiv e-prints

\bibitem[{{Roncarelli} {et~al}\mbox{.}(2007){Roncarelli}, {Moscardini},
  {Borgani}, \& {Dolag}}]{2007MNRAS.378.1259R}
{Roncarelli} M., {Moscardini} L., {Borgani} S., {Dolag} K., 2007, \mnras, 378,
  1259

\bibitem[{{Sadeh} \& {Rephaeli}(2004)}]{2004NewA....9..373S}
{Sadeh} S., {Rephaeli} Y., 2004, \na, 9, 373

\bibitem[{{Sadeh}, {Rephaeli} \& {Silk}(2007){Sadeh}, {Rephaeli}, \&
  {Silk}}]{2007MNRAS.380..637S}
{Sadeh} S., {Rephaeli} Y., {Silk} J., 2007, \mnras, 380, 637

\bibitem[{{Sch{\"a}fer} {et~al}\mbox{.}(2006){Sch{\"a}fer}, {Pfrommer},
  {Bartelmann}, {Springel}, \& {Hernquist}}]{2006MNRAS.370.1309S}
{Sch{\"a}fer} B.~M., {Pfrommer} C., {Bartelmann} M., {Springel} V., {Hernquist}
  L., 2006, \mnras, 370, 1309

\bibitem[{{Schmidt} \& {Allen}(2007)}]{2007MNRAS.379..209S}
{Schmidt} R.~W., {Allen} S.~W., 2007, \mnras, 379, 209

\bibitem[{{Sehgal} {et~al}\mbox{.}(2010){Sehgal}, {Bode}, {Das},
  {Hernandez-Monteagudo}, {Huffenberger}, {Lin}, {Ostriker}, \&
  {Trac}}]{2010ApJ...709..920S}
{Sehgal} N., {Bode} P., {Das} S., {Hernandez-Monteagudo} C., {Huffenberger} K.,
  {Lin} Y.-T., {Ostriker} J.~P., {Trac} H., 2010, \apj, 709, 920

\bibitem[{{Seljak}, {Burwell} \& {Pen}(2001){Seljak}, {Burwell}, \&
  {Pen}}]{2001PhRvD..63f3001S}
{Seljak} U., {Burwell} J., {Pen} U.-L., 2001, \prd, 63, 063001

\bibitem[{{Shaw} {et~al}\mbox{.}(2010){Shaw}, {Nagai}, {Bhattacharya}, \&
  {Lau}}]{2010ApJ...725.1452S}
{Shaw} L.~D., {Nagai} D., {Bhattacharya} S., {Lau} E.~T., 2010, \apj, 725, 1452

\bibitem[{{Shaw} {et~al}\mbox{.}(2009){Shaw}, {Zahn}, {Holder}, \&
  {Dor{\'e}}}]{2009ApJ...702..368S}
{Shaw} L.~D., {Zahn} O., {Holder} G.~P., {Dor{\'e}} O., 2009, \apj, 702, 368

\bibitem[{{Sheth} \& {Tormen}(1999)}]{1999MNRAS.308..119S}
{Sheth} R.~K., {Tormen} G., 1999, \mnras, 308, 119

\bibitem[{{Shimon}, {Sadeh} \& {Rephaeli}(2011){Shimon}, {Sadeh}, \&
  {Rephaeli}}]{2011MNRAS.412.1895S}
{Shimon} M., {Sadeh} S., {Rephaeli} Y., 2011, \mnras, 412, 1895

\bibitem[{{Shirokoff} {et~al}\mbox{.}(2011){Shirokoff}, {Reichardt}, {Shaw},
  {Millea}, {Ade}, {Aird}, {Benson}, {Bleem}, {Carlstrom}, {Chang}, {Cho},
  {Crawford}, {Crites}, {de Haan}, {Dobbs}, {Dudley}, {George}, {Halverson},
  {Holder}, {Holzapfel}, {Hrubes}, {Joy}, {Keisler}, {Knox}, {Lee}, {Leitch},
  {Lueker}, {Luong-Van}, {McMahon}, {Mehl}, {Meyer}, {Mohr}, {Montroy},
  {Padin}, {Plagge}, {Pryke}, {Ruhl}, {Schaffer}, {Spieler}, {Staniszewski},
  {Stark}, {Story}, {Vanderlinde}, {Vieira}, {Williamson}, \&
  {Zahn}}]{2011ApJ...736...61S}
{Shirokoff} E. {et~al.}, 2011, \apj, 736, 61

\bibitem[{{Springel}, {White} \& {Hernquist}(2001){Springel}, {White}, \&
  {Hernquist}}]{2001ApJ...549..681S}
{Springel} V., {White} M., {Hernquist} L., 2001, \apj, 549, 681

\bibitem[{{Springel} {et~al}\mbox{.}(2005){Springel}, {White}, {Jenkins},
  {Frenk}, {Yoshida}, {Gao}, {Navarro}, {Thacker}, {Croton}, {Helly},
  {Peacock}, {Cole}, {Thomas}, {Couchman}, {Evrard}, {Colberg}, \&
  {Pearce}}]{2005Natur.435..629S}
{Springel} V. {et~al.}, 2005, \nat, 435, 629

\bibitem[{{Stanek}, {Rudd} \& {Evrard}(2009){Stanek}, {Rudd}, \&
  {Evrard}}]{2009MNRAS.394L..11S}
{Stanek} R., {Rudd} D., {Evrard} A.~E., 2009, \mnras, 394, L11

\bibitem[{{Sun} {et~al}\mbox{.}(2009){Sun}, {Voit}, {Donahue}, {Jones},
  {Forman}, \& {Vikhlinin}}]{2009ApJ...693.1142S}
{Sun} M., {Voit} G.~M., {Donahue} M., {Jones} C., {Forman} W., {Vikhlinin} A.,
  2009, \apj, 693, 1142

\bibitem[{{Tinker} {et~al}\mbox{.}(2008){Tinker}, {Kravtsov}, {Klypin},
  {Abazajian}, {Warren}, {Yepes}, {Gottl{\"o}ber}, \&
  {Holz}}]{2008ApJ...688..709T}
{Tinker} J., {Kravtsov} A.~V., {Klypin} A., {Abazajian} K., {Warren} M.,
  {Yepes} G., {Gottl{\"o}ber} S., {Holz} D.~E., 2008, \apj, 688, 709

\bibitem[{{Trac}, {Bode} \& {Ostriker}(2011){Trac}, {Bode}, \&
  {Ostriker}}]{2011ApJ...727...94T}
{Trac} H., {Bode} P., {Ostriker} J.~P., 2011, \apj, 727, 94

\bibitem[{{Vikhlinin} {et~al}\mbox{.}(2009){Vikhlinin}, {Burenin}, {Ebeling},
  {Forman}, {Hornstrup}, {Jones}, {Kravtsov}, {Murray}, {Nagai}, {Quintana}, \&
  {Voevodkin}}]{2009ApJ...692.1033V}
{Vikhlinin} A. {et~al.}, 2009, \apj, 692, 1033

\bibitem[{{Vikhlinin} {et~al}\mbox{.}(2006){Vikhlinin}, {Kravtsov}, {Forman},
  {Jones}, {Markevitch}, {Murray}, \& {Van Speybroeck}}]{2006ApJ...640..691V}
{Vikhlinin} A., {Kravtsov} A., {Forman} W., {Jones} C., {Markevitch} M.,
  {Murray} S.~S., {Van Speybroeck} L., 2006, \apj, 640, 691

\bibitem[{{Wang} {et~al}\mbox{.}(2004){Wang}, {Khoury}, {Haiman}, \&
  {May}}]{2004PhRvD..70l3008W}
{Wang} S., {Khoury} J., {Haiman} Z., {May} M., 2004, \prd, 70, 123008

\bibitem[{{Wojtak} \& {{\L}okas}(2010)}]{2010MNRAS.408.2442W}
{Wojtak} R., {{\L}okas} E.~L., 2010, \mnras, 408, 2442

\bibitem[{{Zitrin} {et~al}\mbox{.}(2011){Zitrin}, {Broadhurst}, {Barkana},
  {Rephaeli}, \& {Ben{\'{\i}}tez}}]{2011MNRAS.410.1939Z}
{Zitrin} A., {Broadhurst} T., {Barkana} R., {Rephaeli} Y., {Ben{\'{\i}}tez} N.,
  2011, \mnras, 410, 1939

\end{thebibliography}
\label{lastpage}
\end{document}